\documentclass[%
reprint,
 amsmath, amssymb,
 aps,
]{revtex4-2}

\usepackage{graphicx}
\usepackage{dcolumn}
\usepackage{bm}
\usepackage{algorithm}
\usepackage{algpseudocode}
\usepackage{todonotes}
\usepackage{placeins}
\usepackage{soul}
\usepackage{appendix}
\usepackage{chngcntr}

\newcommand{\bigpalpha}{\mesoP^\alpha}
\newcommand{\dt}{\Delta t}
\newcommand{\mesoP}{\Pi}
\newcommand{\microP}{\pi}
\newcommand{\meso}{\mathbf{T}}
\newcommand{\micro}{\mathbf{P}}
\newcommand{\lag}{\lambda}
\newcommand{\palpha}{\microP^\alpha}
\newcommand{\ppalpha}{\micro^\alpha}
\newcommand{\ppbeta}{\micro^\beta}
\newcommand{\mesom}{\mesoP_m}
\newcommand{\meson}{\mesoP_n}
\newcommand{\microi}{\microP_i}
\newcommand{\microij}{\micro_{ij}}
\newcommand{\qmeso}{Q}
\newcommand{\qmicro}{q}
\newcommand{\tab}{\meso^{\mathrm{abs}}}
\newcommand{\tmn}{\meso_{mn}}
\newcommand{\talpha}{\meso^\alpha}
\newcommand{\tbeta}{\meso^\beta}
\newcommand{\wbar}{\bar{w}}
\newcommand{\wvec}{w}

\let\OLDthebibliography\thebibliography
\renewcommand\thebibliography[1]{
  \OLDthebibliography{#1}
  \setlength{\parskip}{2pt}
  \setlength{\itemsep}{2pt plus 0.3ex}
}

\begin{document}

\preprint{}

\title{Unbiased estimation of equilibrium, rates, and committors from Markov state model analysis}
\author{John D. Russo}
\author{Jeremy Copperman}%
\author{Daniel M. Zuckerman}
 \email{zuckermd@ohsu.edu}
\affiliation{%
 Department of Biomedical Engineering,
Oregon Health and Science University, Portland, OR
}

\author{David Aristoff}
\affiliation{Department of Mathematics, Colorado State University, Fort Collins, CO}
\author{Gideon Simpson}
\affiliation{Department of Mathematics, Drexel University, Philadelphia, PA}

\date{\today}

\begin{abstract}
Markov state models (MSMs) have been broadly adopted for analyzing molecular dynamics trajectories, but the approximate nature of the models that results from coarse-graining into discrete states is a long-known limitation.  We show theoretically that, despite the coarse graining, in principle MSM-like analysis can yield unbiased estimation of key observables.  We describe unbiased estimators for equilibrium state populations, for the mean first-passage time (MFPT) of an arbitrary process, and for state committors – i.e., splitting probabilities.  Generically, the estimators are only asymptotically unbiased but we describe how extension of a recently proposed reweighting scheme can accelerate relaxation to unbiased values.  Exactly accounting for ‘sliding window’ averaging over finite-length trajectories is a key, novel element of our analysis.  
In general, our analysis indicates that coarse-grained MSMs are  asymptotically unbiased for steady-state properties only when appropriate boundary conditions (e.g., source-sink for MFPT estimation) are applied directly to trajectories, prior to calculation of the appropriate transition matrix.
\end{abstract}

\maketitle


\section{Introduction}
Despite modern capabilities to routinely generate multi-microsecond datasets of molecular dynamics (MD) simulation trajectories, the analysis of such trajectories has a notable challenge.  A widespread approach, and the focus of our study, is the Markov state model (MSM) framework which coarse-grains continuous molecular configuration space into discrete states, followed by construction of an approximate transition (stochastic) matrix at a finite lag time from which observables are then calculated \cite{CNrev, pande-msmconformation, pande-noe-book, noe-molecularkinetics, CSP, pande-art}.  The MSM framework, with variations, has also proved useful in analyzing data from rare-events sampling methods, such as the weighted ensemble approach \cite{JACSfold, haMSM, dixon-msm}.
Even as the field has developed more sophisticated MSM analyses \cite{OOM, noe-molecularkinetics, noe-projected, noe-spectroscopy, noe-reversible, noe-msmmilestoning, noe-augmented, noe-bayesian}, recent work has highlighted the approximate nature of MSM results, which results from intrinsic coarse-graining in space and time \cite{CNrev}.  One study highlighted variations of MSM analysis which yielded divergent estimates for multiple observables \cite{Voelz}.  A second report \cite{NCZ} showed that MSMs yield accurate mean first-passage time (MFPT) estimates only for fairly long ($\sim$ 100 ns) lag times for protein folding systems, but also that MSMs are inaccurate for mechanistic characterization which typically reflects shorter timescale behavior. On the other hand, the same analysis showed that including history information – tracing back trajectories to macrostates of interest – enabled accurate MFPT and mechanism characterization within a MSM-like formulation.  These findings largely motivate the present report, where the use of history information is essentially recast as appropriate use of boundary conditions.
Our work is also related to ideas integral to exact milestoning \cite{bello2015exact,aristoff2016mathematical} and non-equilibrium umbrella sampling \cite{warmflash2007umbrella,dickson2009nonequilibrium}.

Our theoretical analysis rests on multiple pillars, several of which appear to be novel.
\begin{itemize}
    \item Most importantly, building on early MSM work \cite{SSP}, we compare ‘microscopic’ discrete-state models with coarse-grained MSM models – of different kinds – which would be generated from trajectories of the microscopic model. In contrast to earlier work, we address the question by exact computation of MSM-like estimates, as opposed to simulating trajectories and thereby introducing the confounding issue of finite sampling. 
    \item Motivated by recent history-augmented MSMs \cite{SAZ, JACSfold, haMSM, NCZ}, we carefully account for boundary conditions (BCs), assessing the difference between applying BCs before or after calculating MSM transition matrices.  Properly accounting for BCs is essential for unbiased estimation of the MFPT and committor. 
    \item We account exactly for initial state bias – i.e., the effects of the expected deviation of trajectories, especially their initial points, from the stationary distribution of interest.  Initial state bias is intrinsic to MSM calculations; if it were not, the required distribution would already be in hand.  
    \item We account exactly for sliding-window averaging occurring over finite-length trajectories.  This averaging is generally used to build MSMs at lag $\tau$ based on examining the pairs of points, $\{(0,\tau), (\dt, \dt+\tau), (2\dt, 2 \dt+\tau), …\}$, where $\dt$ is the spacing between MD trajectory frames. Sliding-window averaging critically underpins – and limits – relaxation to unbiased observable values.  
    \item Finally, we extend a recent proposal to reweight trajectories based on initial estimates of equilibrium distributions \cite{Voelz}, by iterating this process to self-consistent convergence and additionally applying it to non-equilibrium stationary conditions.  Our analysis shows that reweighting can significantly reduce the relaxation time required to achieve unbiased estimates of observables.
\end{itemize}

\section{Theoretical framework}

\subsection{Notation}

For clarity, we define the various symbols used throughout this work in Table~\ref{tab:symbol-definitions}.

\begin{table}[h]
    \caption{Definitions of symbols used in this work.}
    \label{tab:symbol-definitions}
    \centering
    \begin{tabular}{c | @{\hspace{1em}}l@{}}
        Symbol & Definition \\ [0.5ex] \hline
        $i, j$ & Microstates (single phase points) \\
        $\micro$ & Microscopic transition matrix \\
        $m, n$ & Coarse states (sets of microstates) \\
        $\meso$ & Coarse-grained transition matrix \\
        $\microP$ & Microscopic equilibrium distribution \\
        $\mesoP$ & Coarse-grained equilibrium distribution \\
        $\micro^\alpha$, $\meso^\alpha$ & A $\rightarrow$ B steady-state matrices \\
        $\microP^\alpha$, $\mesoP^\alpha$ & A $\rightarrow$ B steady-state distributions \\
        $w$ & Microstate weights \\
        $\wbar$ & Sliding-window averaged microstate weights \\
        $S$ & Trajectory length: number of steps \\
        $\dt$ & Timestep of microscopic model\\
        $\tau$ & Physical lag time \\
        $\lambda$ & Dimensionless lag time, $\tau/\dt$ \\
        $q$ & Microscopic committor to state A \\
        $Q$ & Coarse-grained committor to state A
    \end{tabular}
\end{table}

\subsection{Fine- and coarse-grained systems}

The process of building a Markov model typically involves simulating continuous trajectories, discretizing them by assigning points in the trajectories to states, and constructing the model on the space of states. \cite{CNrev}
Taking the continuous Markovian phase space described by the system's microscopic dynamics and grouping it into discrete states produces states that cannot be perfectly Markovian. \cite{OOM, NCZ, SAZ}

Examining the effects of coarse-graining on trajectories is complicated by the sampling issues present in any trajectory analysis.
We therefore employ a framework for exactly recapitulating the process of constructing a coarse-grained model from trajectories, without using actual trajectories.
This enables us to study the coarse-graining exactly, without any sampling concerns.

For simplicity, we use a discrete representation for the underlying dynamics, although as we discuss, our results are expected to apply for continuous dynamics as well.
Let $\micro$ be the underlying fine-grained, Markovian transition matrix for a single time step $\Delta t$. 
For simplicity we assume $\micro$ is a finite matrix.
Note that we use the term \emph{microstate} in its traditional statistical mechanics sense to connote a single phase-space point or discrete state; this contrasts with the ambiguous usage of the term in the MSM community to describe a coarse-grained state. \cite{pande-noe-book,CNrev}
The matrix $\micro$ will implicitly account for boundary conditions (BCs) chosen according to the observable of interest.
Here we refer to boundary conditions applied between two macrostates A and B of interest -- i.e., source-sink BCs, dual-absorbing BCs, and the absence of sources or sinks.
The issue of boundary conditions is central to our analysis and will be described in further detail below.

The coarse-grained MSM transition matrix $\meso$ is obtained by merging microstates of the fine-grained model.
The resulting transition probability from coarse state $m$ to any other coarse state $n$ will be a weighted average over microscopic transition probabilities:
\begin{equation}
    \meso_{m \rightarrow n}(S) = \left.
    \sum_{i \in m} \sum_{j \in n} 
    \wbar_i(S) \,
    \micro_{i \rightarrow j}
    \right/ \sum_{i \in m} \wbar_i(S) \; .
    \label{eq:coarse-graining}
\end{equation}
The microstate weights $\wbar_i$ are computed to exactly mimic  the process of counting transitions in $S$-step trajectories but without sampling error, as described below.
Also note that the coarse-grained MSM transition matrix $\meso$ will ``inherit'' the BCs of the matrix $\micro$ as described below.

\subsection{Accounting for finite trajectory length in sliding window averaging}

To compute the necessary averages for the MSM transition matrix $\meso(S)$, we must account both for the initial distribution of trajectories as well as the subsequent dynamics and relaxation that occurs.
To do so, we let $w_i(t)$ be the time-evolving weight of microstate $i$, which represents the fraction of trajectories in state $i$ at time $t$.
The set of weights is assumed to be normalized over the full microscopic space, so that
\begin{equation}
    \label{eq:wnorm}
    \sum_i w_i(t) = 1 \; .
\end{equation}

Once the set of $w_i(0)$ is defined, the time evolution of this distribution is fully determined by the underlying transition matrix $\micro$ according to
\begin{equation}
    \wvec(t+\Delta t) = \wvec(t) \, \micro
    \label{eq:weightt}
\end{equation}
where $\wvec$ is the vector of weights $w_i$.
Importantly, we do not generate trajectories, and there are no sampling limitations in our analysis.
Instead our calculations yield the same results as if there were an \textit{infinite number} of \textit{finite-length} trajectories.

Trajectories are taken to consist of $S$ steps or $S+1$ time points indexed by $\{0, 1, 2, ..., S\}$.

We can now replicate the sliding-window average used in MSM construction \cite{CNrev} by averaging over the time evolving distribution.
The time-averaged weights for a single-step lag time are given by 
\begin{equation}
    \wbar_i(S) = \frac{1}{S} \sum_{s=0}^{S-1} w_i(s \Delta t), 
    \label{eq:time-avg-weights}
\end{equation}
where  $\wvec(s \Delta t) = \wvec(0) \, \micro^s$ is the weight distribution as evolved according to the Markovian microscopic model $\micro$.
These time-averaged weights are normalized because the instantaneous weights sum to one.

Note that if the initial weights are not in the stationary distribution of interest -- e.g., equilibrium or a non-equilibrium steady-state (NESS) -- then we expect the corresponding estimates for $\meso$ in Eq.~\ref{eq:coarse-graining} to be biased, unless trajectories are much longer than the associated relaxation process.

\subsubsection*{Generalization to arbitrary lag time}

The sliding window calculation can be generalized to arbitrary lag time $\lag = \tau/\dt > 1$, where $\tau$ is the physical lag time.
The window starts at the first point in the trajectory ($s=0$), and ends $\lag$ steps from the end of the trajectory.
Less data is averaged because the final steps are omitted as start points of the window, but more relaxation occurs compared to $\lag=1$.
Eq.~\ref{eq:time-avg-weights} becomes
\begin{equation}
   \wbar_i(S, \lag) = \frac{1}{S - \lag + 1} \sum_{s=0}^{S-\lag} w_i(s \Delta t)
   \label{eq:time-avgd-weights-arb-lag}
\end{equation}
where the individual weights are again determined by \eqref{eq:weightt}.
Note that the lag time, which is used only for analysis, does not affect the underlying dynamics embodied in $\micro$ and $\wvec(t)$.
With this, we can write the arbitrary lag time coarse-grained transition matrix as
\begin{equation}
\left .
    \meso_{m \rightarrow n} (S, \lag) = 
    \sum_{i \in m} \sum_{j \in n} 
     \wbar_i(S, \lag) \,
    \micro_{i \rightarrow j}^\lag
    \middle/ \sum_{i \in m} \wbar_i(S, \lag)
\right .
    .
    \label{eq:time-bias-cg-weighted-lag}
\end{equation}

For clarity of presentation, the rest of this work uses $\lag = 1$, though analogous results apply to any $\lag$.

Also note that we can restrict averaging in \eqref{eq:time-avg-weights} and \eqref{eq:time-avgd-weights-arb-lag} to later time points in the trajectories (i.e., start the sums at $s>0$), which will exclude earlier, less relaxed time points.
This will be explored in subsequent work.

\subsection{Accounting for boundary conditions}

To our knowledge, the issue of boundary conditions (BCs) has not been addressed thoroughly in the MSM literature.
BCs are fundamental to MSM construction because the transition matrix is determined by the average \emph{intra}-coarse state distribution $\wbar$ as seen in \eqref{eq:coarse-graining} and \eqref{eq:time-bias-cg-weighted-lag}, which in turn depends on other coarse states because of the time evolution of the distribution \eqref{eq:weightt} -- i.e., on transitions between coarse states which are constrained by the BCs.
From this perspective, it is not surprising that the (equilibrium-like) lack of BCs will lead to unbiased equilibrium populations and that source-sink BCs will lead to unbiased MFPTs.
The situation for committors is essentially a hybrid of the two as explained below.

As our data will show, failure to account for BCs correctly can lead to biased estimators.
For example, computing a first-passage time involves measuring the time from when trajectories enter (or are initiated in) some state A to when they first enter another state B, including any returns to state A.
Such trajectories are consistent with a sink at B and a source at A, i.e., source-sink (``recycling'') boundary conditions, 
as required for computing the MFPT via the Hill relation \eqref{eq:mfpt-micro}
given below.
However, if trajectories were allowed to emerge from the sink state B and re-enter B without first returning to A, such events would bias MFPT estimation using a MSM transition matrix.
Asymptotically, the intra-coarse state distributions $\wbar$ would not match the NESS and hence the transition matrices would not be appropriate for unbiased MFPT computation.

The committor describes a dually absorbing process at two states A and B.
We let $\qmicro_i$ be the committor to A, the fraction of trajectories absorbed to A starting from microstate $i$; the committor to B is $1-\qmicro_i$.
If absorbing conditions at A and B are not enforced in the microscopic model (the trajectories), we expect committor estimates to be biased, even for coarse states which consist of collections of microstates.
However, as will be seen, simply building a transition matrix from dually absorbing trajectories is not a route to unbiased committors.

Equilibrium, on the other hand, requires detailed balance, so no sources or sinks can be present.
Correspondingly, equilibrium probabilities are computable without bias, asymptotically, from a standard MSM. 

Below we will consider several types of coarse transition matrices (MSMs) built from different boundary conditions embodied in the microscopic transition matrix.
All MSMs are constructed from the same weight formulation, namely \eqref{eq:coarse-graining} for single-step lag  ($\lag=1$) or \eqref{eq:time-bias-cg-weighted-lag} for $\lag>1$, using the microscopic transition matrix $\micro$ corresponding to  different boundary conditions as follows: 
\begin{itemize}
\item The standard MSM denoted $\meso$ is derived from the full microscopic model $\micro$ with no boundary conditions applied.
\item A source-sink ssMSM can be constructed for either the A $\to$ B direction, denoted $\talpha$, or the B $\to$ A direction, called $\tbeta$. The matrix $\talpha$ is constructed from the modified microscopic model $\ppalpha$ which is obtained from $\micro$ by setting $\micro_{ij} = 0$ for $i \in B$ except when $j \in A$.
For simplicity, here we assume states A and B each consist of a single microstate, so that $\micro_{BA} = 1$ for this ssMSM.
The coarse matrix $\tbeta$ is constructed in analogous fashion from $\ppbeta$.

\item The dually absorbing abMSM denoted $\tab$ is obtained by setting $\micro_{ij} = 0$ for $i \in$ A \emph{or} B.  The abMSM, although it seems natural for computing committors, will be seen to be biased.
\end{itemize}

\subsection{Unbiased asymptotic estimators}
\label{sec:proofs}

We now demonstrate that Markov-like models with the appropriate boundary conditions incorporated into the trajectories during construction produce unbiased estimates of equilibrium probabilities for coarse-grained states, the first-passage time, and the set of coarse-grained committors.
Our argument relies on two simple parts.
First, we note that under boundary conditions allowing for stationarity, regardless of the initial weights, the average weights $\wbar_i$ asymptotically approach their stationary values.
Second, we show that the stationary weights (reached asymptotically) yield unbiased observables with appropriate estimators.

The asymptotic stationarity of the time-averaged weights $\wbar$, defined by \eqref{eq:time-avg-weights} or \eqref{eq:time-avgd-weights-arb-lag}, follows from the fact that the weights constitute an ordinary probability distribution in the microscopic space evolving under standard Makrovian dynamics \eqref{eq:weightt}. We will assume that $\micro$ 
is irreducible, meaning all regions in the microscopic state 
space are connected by positive probability 
paths, \cite{norris1998markov}. 
With this assumption, recalling that $\micro$ in \eqref{eq:weightt} is assumed to embody any boundary conditions, 
we see that under equilibrium or source-sink ($\alpha$) conditions, 
the time-averaged weights will 
approach $\microP$ or $\microP^\alpha$, correspondingly,
\begin{equation}
    \wbar \to \microP 
    \hspace{0.5cm}\mbox{ or } \hspace{0.5cm}
    \wbar \to \microP^\alpha,
    \label{eq:wbar-asymp}
\end{equation}
under equilibrium or $\alpha$ source-sink BCs as $S \to \infty$.

It will prove convenient to show a related result, namely, that coarse-grained stationary probabilities $\mesoP$ derived using the  stationary weights are exactly the sums of the corresponding microscopic stationary probabilities $\microP$.
Starting from the coarse stationarity condition, we use the asymptotic stationary weights \eqref{eq:wbar-asymp} along with the coarse matrix \eqref{eq:coarse-graining} to find
\begin{align}
    \meson &= \sum_m \mesom \tmn(S \to \infty) \\
    &= \left. \sum_m \mesom \sum_{i\in m} \sum_{j\in n} \microi \microij \right/ \sum_{i\in m} \microi \;.
\end{align}
If we substitute $\mesom = \sum_{i \in m} \microi$ 
into the right-hand side of this expression, we find
\begin{align}
    \meson &= \sum_m  \sum_{i\in m} \sum_{j\in n} \microi \microij \\
    &= \sum_{j \in n} \sum_i \microi \microij \\
    &= \sum_{j \in n} \microP_j \; ,
    \label{eq:cg-sum}
\end{align}
which demonstrates the consistency of the summed microscopic stationary probabilities with coarse-grained stationarity. 
This completes the demonstration.

Note that the result \eqref{eq:cg-sum} holds regardless of boundary conditions, so long as the stationary probabilities and transition matrix are for the same BCs.  In particular, it implies
\begin{equation}
    \mesoP^\alpha_n = \sum_{j \in n} \microP^\alpha_j
    \label{eq:cg-sum-alpha}
\end{equation}
for the $\alpha$ (A to B) NESS.

We now consider the different observables in turn and show that asymptotically, when the average weights approach stationary values, suitable coarse-grained estimators become unbiased.
That is, we must show that estimators obtained solely from calculations using coarse-grained $\meso$ matrices asymptotically yield observables in exact agreement with microscopic values.

\subsubsection{Equilibrium}

Coarse-grained equilibrium probabilities $\mesoP$ can be estimated without bias as the stationary solution to the standard MSM in the limit of infinite trajectory length:
\begin{equation}
    \mesoP \, \meso(S \to \infty) = \mesoP
    \label{eq:stationary}
\end{equation}
This follows from the asymptotic stationarity of the weights \eqref{eq:wbar-asymp}, which in turn causes the coarse-grained stationary probabilities to match the sum of microscopic stationary probabilities as in \eqref{eq:cg-sum}. It is easy to check 
that the conditions above on $\micro$ ensure 
that $\meso(S)$ has a unique stationary 
distribution for large enough $S$.

\subsubsection{Mean first-passage time}

We employ a similar strategy for the MFPT, showing that macroscopic analog of the microscopic solution recapitulates the microscopic value, \emph{so long as the correct source-sink boundary conditions are employed.}
We make use of the Hill relation, which relates the source-sink steady-state flux into a target macrostate $B$ to the MFPT(A $\to$ B) according to~\cite{hill}
\begin{equation}
    1/\textrm{MFPT} = \mathrm{Flux} (A \rightarrow B) \; .
\end{equation}
Recalling that the A $\to$ B NESS is designated by $\alpha$, we recast the flux using the microscopic model to yield the reference dimensionless expression
\begin{equation}
    \dt/\textrm{MFPT} = \sum_{i \notin B} \sum_{j \in B} \microP_i^\alpha \micro^\alpha_{ij} \; .
    \label{eq:mfpt-micro}
\end{equation}
We will explore corase-grained estimates of the MFPT generically given by the analogous expression
\begin{equation}
    \dt/\textrm{MFPT} 
    = \sum_{m \notin B} \sum_{n \in B} \mesoP_m \meso_{mn}(S) \; .
    \label{eq:mfpt-cg}
\end{equation}

We now show that using the flux computed from the asymptotic coarse ssMSM yields a MFPT identical to that from the microscopic model.
Using the $\alpha$-specific asymptotic weights \eqref{eq:wbar-asymp} in the coarse ssMSM $\talpha$ defined by \eqref{eq:coarse-graining}, we obtain
\begin{align}
    \dt/\textrm{MFPT} 
    &= \sum_{m \notin B} \sum_{n \in B} \mesoP_m^\alpha \meso^\alpha_{mn}(S \to \infty) \\
    &= \sum_{m \notin B} \sum_{n \in B} \mesoP_m^\alpha 
    \left(
    \left.
    \sum_{i \in m} \sum_{j \in n} \microP^\alpha_i \micro^\alpha_{ij} \middle/ \sum_{i \in m} \microP^\alpha_i 
    \right.
    \right) \\
    &= \sum_{m \notin B} \sum_{n \in B} \sum_{i \in m} \sum_{j \in n} \palpha_i \micro^\alpha_{ij} \\
    &= \sum_{i \notin B} \sum_{j \in B} \palpha_i \micro^\alpha_{ij} \; ,
\end{align}
where we made use of \eqref{eq:cg-sum-alpha}.
Hence the MFPT calculated from the ssMSM with asymptotic weights yields the correct microscopic value \eqref{eq:mfpt-micro}. 

We note that our formulation here, including for the microscopic model, retains a discretization error, expected to be $O(\dt/\mathrm{MFPT})$.
This is because $\micro^\alpha_{ij} > 0$ for $j \in B$ will lead to non-zero occupancy of B, with expected probability in B of $\sum_{i \in B} \microP^\alpha_i \sim \dt/\mathrm{MFPT}$ from the definitions of the MFPT and NESS.
Even this small error can be avoided with a slightly more complex formulation, as we will show in future work.

\subsubsection{Committors}

We now demonstrate a novel estimator for coarse-grained committors based on the ratio of the steady-state to equilibrium probabilities.
It has been shown previously that, microscopically, the committor to A, $q$, is proportional to the ratio of the $\alpha$ NESS to equilibrium probabilities~\cite{darve}:
\begin{equation}
    \palpha_i = c \, q_i \microP_i \;,
    \label{eq:qmicro}
\end{equation}
where $c = \palpha_i / \microP_i > 1$ for $i \in$ A. 
We propose to estimate coarse-grained committors $\qmeso$ according to 
\begin{equation}
    \qmeso_m = \bigpalpha_m / c \, \mesoP_m \; ,
    \label{eq:qbig}
\end{equation}
where $c = \bigpalpha_m / \mesoP_m$ for $m \in$ A has the same value as in the microscopic case because of the relations \eqref{eq:cg-sum} and \eqref{eq:cg-sum-alpha}.

It is not immediately obvious what the ``exact'' coarse-grained $\qmeso$ values should be.
Consider a thought-experiment of computing committors from an extremely long `equilibrium' trajectory which traces back and forth between states A and B many times, visiting all microstates.
We could estimate the committor for a coarse state $m$ by considering all time points of the trajectory in $m$ and counting the fraction of downstream trajectory segments which reach A before B for each such time point.
The configurations in the coarse state will be equilibrium distributed due to the length of the trajectory, 
and the fractional absorptions to A and B for segments visiting a given microstate $i \in m$ will necessarily be determined by the microscopic committor $q_i$.
This scenario motivates equilibrium weighting of microscopic committors according to
\begin{equation}
    \qmeso_m = 
    \left . 
    \sum_{i \in m} \microP_i q_i 
    \middle / \sum_{i \in m} \microP_i
    \right . .
    \label{eq:mesocomm-from-micro}
\end{equation}
Indeed, it would be difficult to motivate other choices, such as a uniform weighting or weighting according to a particular directional NESS.

To validate the estimator \eqref{eq:qbig} asymptotically as $S \to \infty$, we substitute the asymptotically exact microscopic decompositions \eqref{eq:cg-sum} and \eqref{eq:cg-sum-alpha} for the coarse stationary probabilities.
This yields
\begin{align}
   \qmeso_m &= 
    \left . 
    \sum_{i \in m} \microP_i^\alpha 
    \middle / c \sum_{i \in m} \microP_i
    \right. 
    \\
    &=
    \left . 
    \sum_{i \in m} c \, q_i \microP_i 
    \middle / c \sum_{i \in m} \microP_i
    \right. ,
\end{align}
where we have used \eqref{eq:qmicro} and recapitulate the desired result \eqref{eq:mesocomm-from-micro}.
Although the suitability of equilibrium weighting among microscopic committors can be debated, the ratio estimator \eqref{eq:qbig} yields this natural average.

\subsection{First-step relation for committors}

As we will see, the abMSM is biased for committor estimates, despite seeming like a natural and correct choice of boundary conditions.
For completeness, we review a procedure for calculating the committor from a transition matrix using a `first-step' relation \cite{apaydin2003stochastic,first-step}.

If the committor to A at a microstate $i$ is given by $q_i$, the average committor of trajectories initiated in that point and propagated for one step is also equal to $q_i$.
The analogous formulation for a coarse model is therefore
\begin{equation}
    \qmeso_m = \sum_{n} \meso_{mn} \qmeso_n 
    \hspace{0.5cm}
    i \notin A, B
    \label{eq:first-step}
\end{equation}
where $\qmeso_{n \in B} = 0$ and $\qmeso_{n \in A} = 1$.
Although not unbiased for coarse states, this relation is used for reference in the results shown below.

\subsection{Iterative reweighting}
\label{sec:reweighting-discuss}

Although we have described estimators that are unbiased asymptotically, deviation in the initial weights $\wbar_i$ from the appropriate steady-state distribution introduces initial-state bias which can be very slow to relax away, as our results will show.
As a trajectory propagates, the relaxation time for the initial distribution to converge to a steady-state distribution will depend on the initial distribution.

Recent work showed that computing steady-state twice, once from an MSM with uniform initial weights for each trajectory, then recalculating the MSM using weighted trajectories (with weights from the first steady-state probability estimate of the initial bin of each trajectory), substantially reduced the trajectory length necessary for converged estimates.~\cite{Voelz}

In fact, this process can be applied iteratively, using the estimate from the previous iteration as the weights for the next.

\begin{algorithm}[H]
\caption{Iterative reweighting algorithm}\label{algo:reweighting}
\begin{algorithmic}[1]
\State Choose uniform initial weights $w_i$
\Repeat
    \State Compute $\bar{w}_i$ from $w_i$ using \eqref{eq:time-avg-weights}
    \State Compute the interim stationary distribution $\tilde \mesoP$ by solving $\tilde \mesoP \meso = \tilde \mesoP$ 
    \State Update the microbin weights $w(0)$ by evenly dividing the coarse probabilities over microbins according to $w_i(0) = \left. \tilde \mesoP_m \middle/ \sum_{j \in m} 1 \right.$ for $i \in m$
\Until desired number of iterations
\end{algorithmic}
\end{algorithm}

Results for iterative reweighting are presented in Sec.~\ref{sec:reweighting}.

\subsection{Connection to continuous trajectories}

We expect that our discrete-state analysis will carry over directly to the case where microscopic dynamics are continuous in space. 
First, one may consider the limit of arbitrarily small microstates, leading to quasi-continuous dynamics.
Second, the derivations presented in Sec.~\ref{sec:proofs} rely almost exclusively on the relaxation of the initial weights to steady-state values, a process will also occur under continuous dynamics.

\section{Results}

Numerical results confirm our theoretical expectations. Equilibrium probabilities, mean first-passage times, and committors of coarse-grained MSMs are unbiased in the asymptotic limit in general only when they are based upon the relaxation of microscopic trajectories to the appropriate steady-state distributions, which can be achieved by sliding window relaxation and applying the appropriate BCs at the microscopic trajectory level.

\begin{figure}
    \centering
    \includegraphics[width=\linewidth]{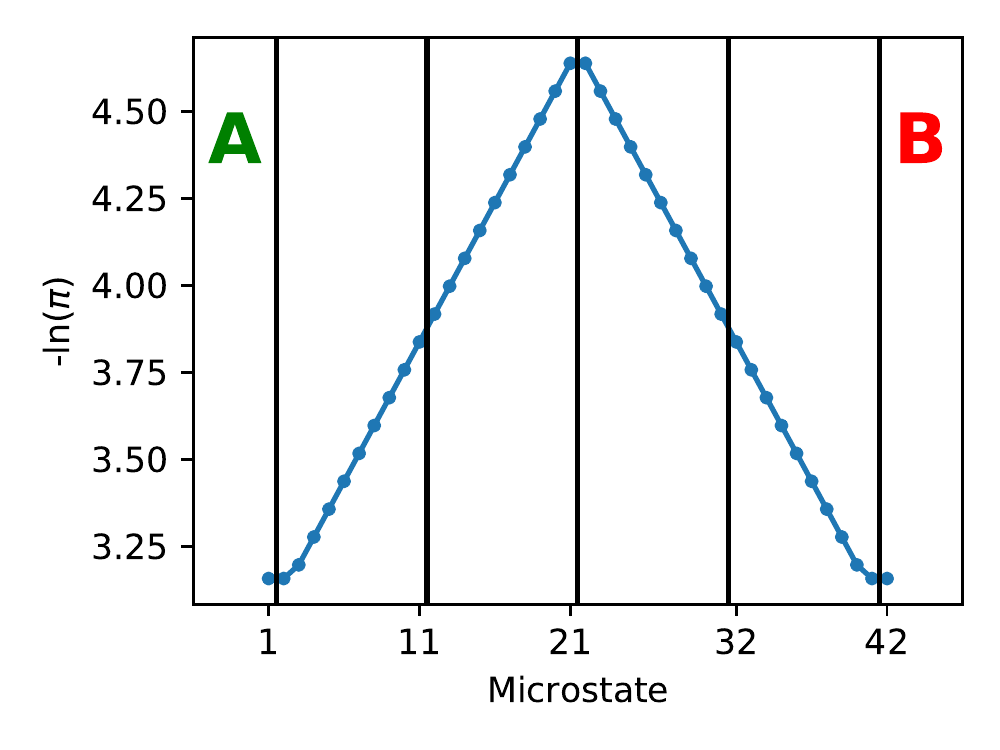}
    \caption{Energy landscape of the 42-microstate fine-grained system.
    Microstate boundaries are denoted by dots.
    The coarse-grained states or `bins' are separated by vertical lines.
    Bins 1-4 are the four intermediate coarse states.
    `Macrostates' A and B, are the leftmost and rightmost individual microstates, respectively, in both fine and coarse descriptions.}
    \label{fig:landscape}
\end{figure}
We demonstrate by estimating equilibrium probabilities, mean first-passage times, and committors on a sample system, where the microscopic dynamics are exactly described by a 42 microstate transition matrix (exact transition probabilities given in Figure \ref{fig:p}). 
The coarse-graining preserves the first and last states as the macrostates A and B, and groups the intermediate 40 microstates into 4 coarse states.
 The energy landscape is shown in Fig.~\ref{fig:landscape}, along with lines indicating the coarse states. The energy landscape of this system emulates two stable states separated by an energy barrier.

This minimal system provides an unambiguous demonstration of how initial state bias affects key observables. In a common procedure, a finite set of trajectories may be generated for MSM construction with initial points spanning the space of interest. We emulate this procedure by introducing uniform initial weights into Eq. \ref{eq:time-avg-weights}, emulating the distribution of trajectory starting points in a finite sample. In this system with a central energy ``barrier'', this constitutes significant initial state bias.
We examine the estimators as a function of trajectory length $S$, to determine both how the initial bias relaxes out with longer trajectories, and what length trajectories are necessary for converged estimates. An iterative approach to accelerate convergence is explored in Sec.~\ref{sec:reweighting}. 
Lag time and trajectory length both contribute to recovering unbiased estimations using trajectories whose initial points are not steady-state distributed.

\FloatBarrier

\subsection{Asymptotic estimators}
\label{sec:asymptotic}

\begin{figure}[h]
    \centering
    \includegraphics[width=0.6\linewidth]{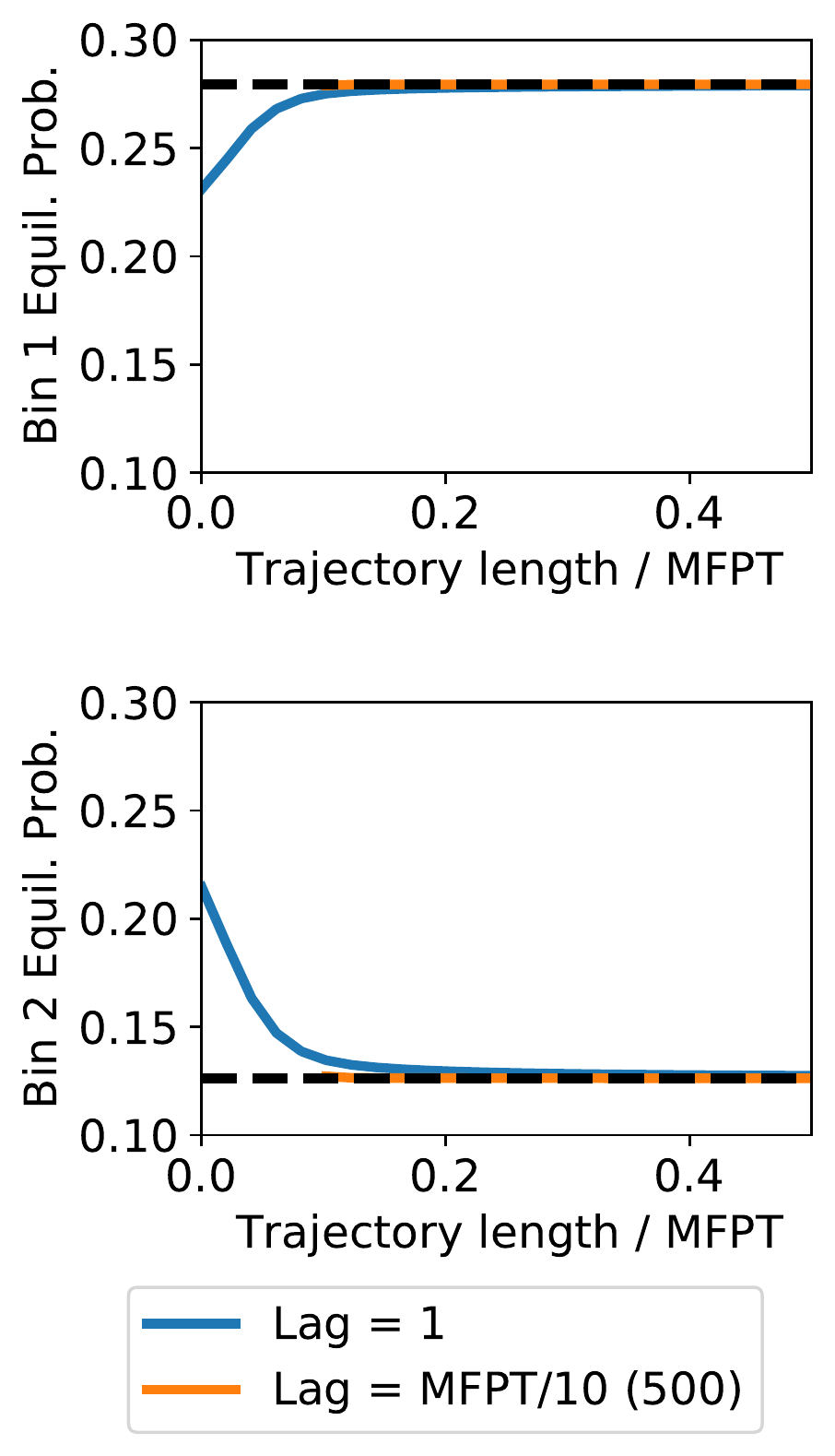}
    \caption{\textbf{MSM equilibrium probability estimates are asymptotically unbiased.} The equilibrium estimator, shown at lag of 1 step ($\lambda=1$, blue) and $\lambda = 500$ (orange). The black dashed line is the exact reference value, computed from the microscopic matrix.
    Because the energy landscape is symmetric,  only states (`bins') in the left half are shown.
    We assumed a non-informative uniform initial distribution of weights $w_i(0) = 1/42$.
    }
    \label{fig:asymp_equil}
\end{figure}

The estimation of equilibrium probabilities is a very straightforward application of a MSM,  and as expected a standard MSM is unbiased both in the limits of long trajectories and long lag times. The reference equilibrium distribution is obtained as the stationary solution of the microscopic transition matrix $\micro$.
We show results for a standard MSM at lag time of 1 and $\mathrm{MFPT}/10=500$ steps in Fig. \ref{fig:asymp_equil}. Note that all results are plotted as a function of the trajectory length, which governs the amount of relaxation that occurs within a trajectory ensemble.
At a lag of 1, the uniform initial weights introduce some initial-state bias, shown in Fig.~\ref{fig:asymp_equil}.
However, this initial bias quickly relaxes out, and converged first-passage time estimates are obtained within $\sim \mathrm{MFPT}/5$ steps.

The longer lag appears to produce estimates closer to the reference values.
However, this is because the minimum trajectory length is given by $\lag + 1$, so the first estimate produced at the longer lag is at a long trajectory length.
At this length, the short-lag estimate was also relaxed to nearly the reference value.

\begin{figure}[h]
    \centering
    \includegraphics[width=0.65\linewidth]{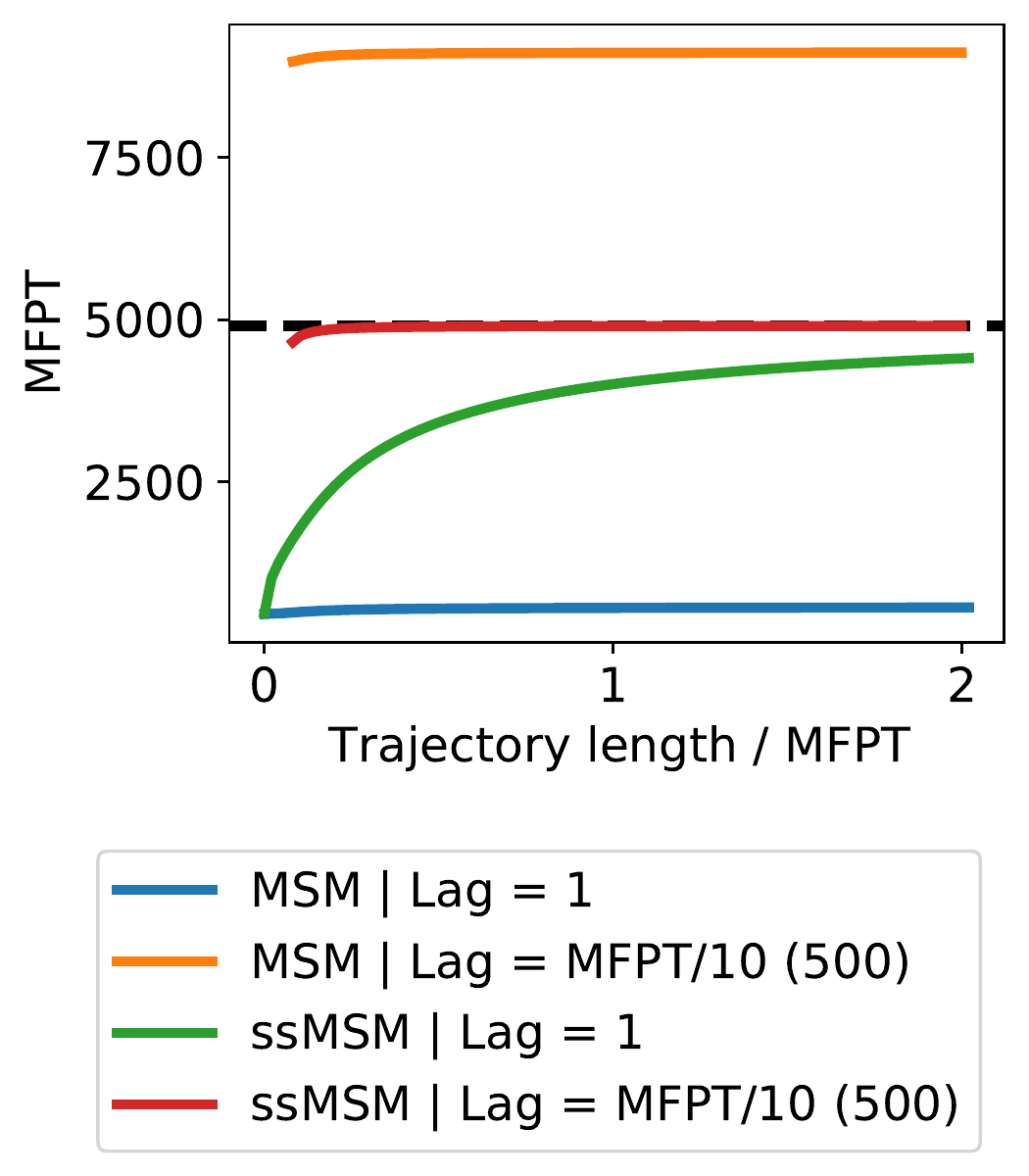}
    \caption{\textbf{Unbiased $\mathrm{MFPT}$ estimation from ssMSMs.} 
    Employing the coarse-grained Hill relation \eqref{eq:mfpt-cg}, we compare MFPT estimates from standard MSMs at short lag time ($\lambda = 1$ step, blue line) and long lag time ($\lambda = 500 \sim \mathrm{MFPT}/10 \dt$, orange line), ssMSM at short lag time ($\lambda=1$, green line) and long lag time $\lambda=500$, red line), and the exact reference value (black dashed line).
    We assumed a non-informative uniform initial distribution of weights $w_i(0) = 1/42$.
    }
    \label{fig:asymp_mfpt}
\end{figure}

Despite the apparent simplicity of this two-state system, first passage times can be significantly biased by the initial state distribution. First-passage times to state $B$ are computed using the Hill relation \eqref{eq:mfpt-micro}, referenced to the MFPT for a lag time of one step ($\dt$, or $\lag = 1$). Here, source-sink boundary conditions are applied to the standard MSM after construction, while for the ssMSM they are applied at the microscopic trajectory level. When the source-sink BCs are not applied at the trajectory level, the $\mathrm{MFPT}$ estimates are significantly biased at lag times of 1 step and $\mathrm{MFPT}/10 \sim 500 \dt$, and do not improve with the trajectory length, shown in Fig.~\ref{fig:asymp_mfpt}. Note that standard MSMs can recapitulate physical MFPTs at long enough lag times\cite{NCZ}. When the BCs are applied (ssMSM), the $\mathrm{MFPT}$ estimate becomes unbiased for trajectories longer than the $\mathrm{MFPT}$ itself. However, combining the application of BCs at the trajectory level (ssMSM), and increasing lag time to $\mathrm{MFPT}/10 \sim 500 \dt$, leads to an unbiased first-passage time estimate at a fraction of the $\mathrm{MFPT}$.

\begin{figure}[h]
    \centering
    \includegraphics[width=\linewidth]{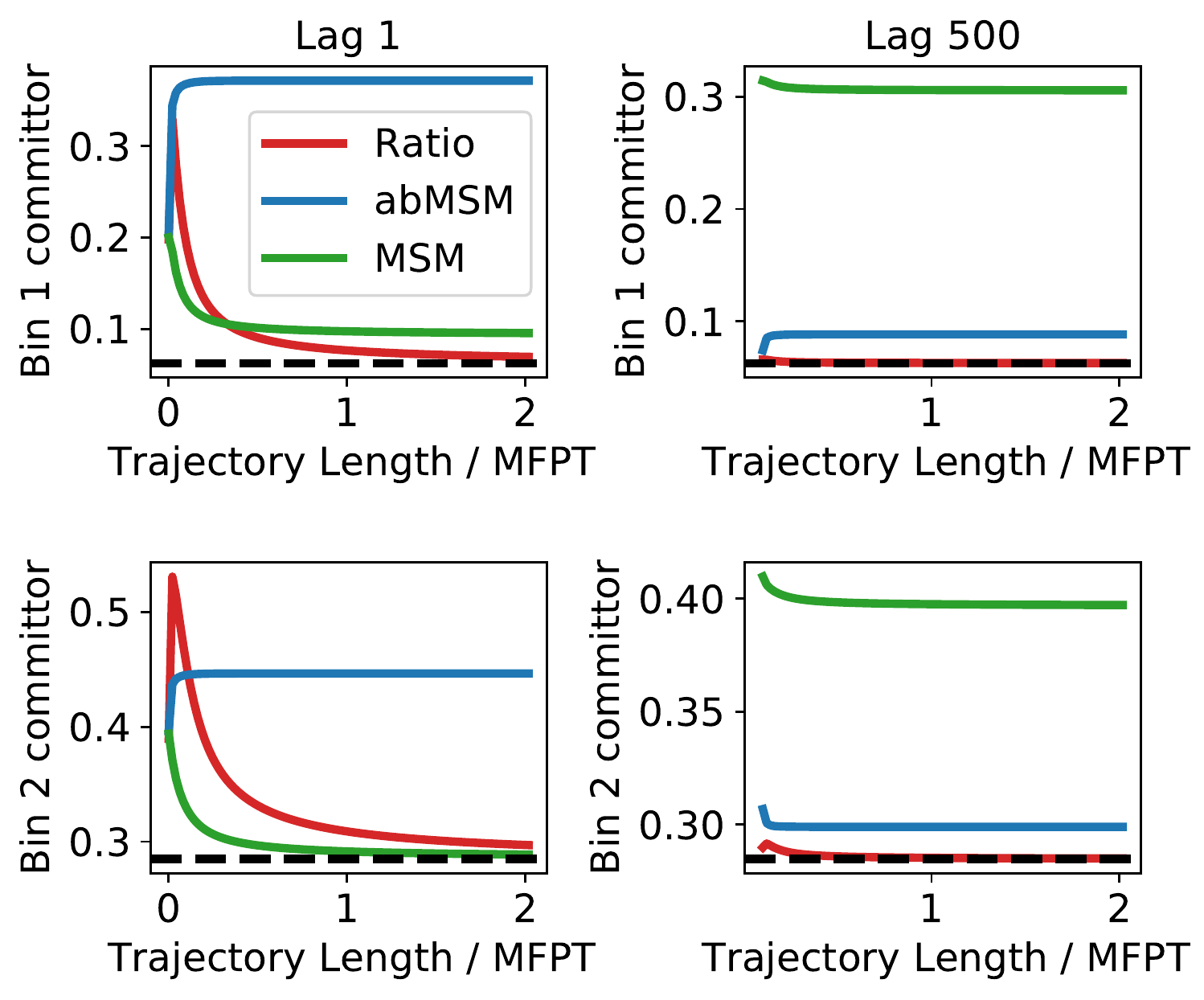}
    \caption{\textbf{Unbiased committor estimation from the steady-state ``ratio method''.}
    Committor estimation using the first-step relation \eqref{eq:first-step} with standard MSM (blue lines) and abMSM (green lines) at short lag time ($\lambda=1$, left) and long lag time $\lambda=500 \sim \mathrm{MFPT}/10\dt$, right), as well as committor estimates from the ratio of equilibrium and NESS (source/sink BCs) steady-states (red lines), compared to the reference value (dashed black line).
    We assumed a non-informative uniform initial distribution of weights $w_i(0) = 1/42$.
    }
    \label{fig:asympt_comm}
\end{figure}

Like the $\mathrm{MFPT}$, the committor stratifying the A to B transition (see Fig. \ref{fig:asympt_comm}) is sensitive to BC application at the trajectory level, and moreover, asymptotically unbiased estimation requires a novel approach. First-step relations \eqref{eq:first-step} applied to the coarse-grained standard MSM estimates are biased at both short and long lag times. Surprisingly, even when appropriate BCs are applied at the trajectory level before MSM construction (i.e., using the abMSM), committor estimates based upon first-step relations are biased at both short and long lag times, even in the limit of long trajectories.
We find that asymptotically unbiased committor estimation requires calculation of the committor via the ratio \eqref{eq:qbig} of the equilibrium and NESS (ssMSM source/sink BCs) steady-state distributions . Since this ``ratio method'' estimator is based upon steady-state distributions, the initial bias can relax and committor estimates converge asymptotically to the reference value. For the longer lag time of $\mathrm{MFPT}/10 \sim 500 \dt$ steps, this relaxation is rapid within a fraction of the $\mathrm{MFPT}$.

\subsection{Iterative reweighting}
\label{sec:reweighting}

The sliding window relaxation time to a steady-state microscopic distribution is a priori unknown, and may be computationally prohibitive. This motivates the exploration of an iterative approach which accelerates steady-state convergence.
By iteratively obtaining estimates of the steady-state and equilibrium distributions, and then using those as the initial weights to compute the estimates again as described in Sec.~\ref{sec:reweighting-discuss}, we can accelerate the relaxation of this initial bias and reduce the trajectory length needed to obtain converged estimates.

\begin{figure}[h]
    \centering
    \includegraphics[width=\linewidth]{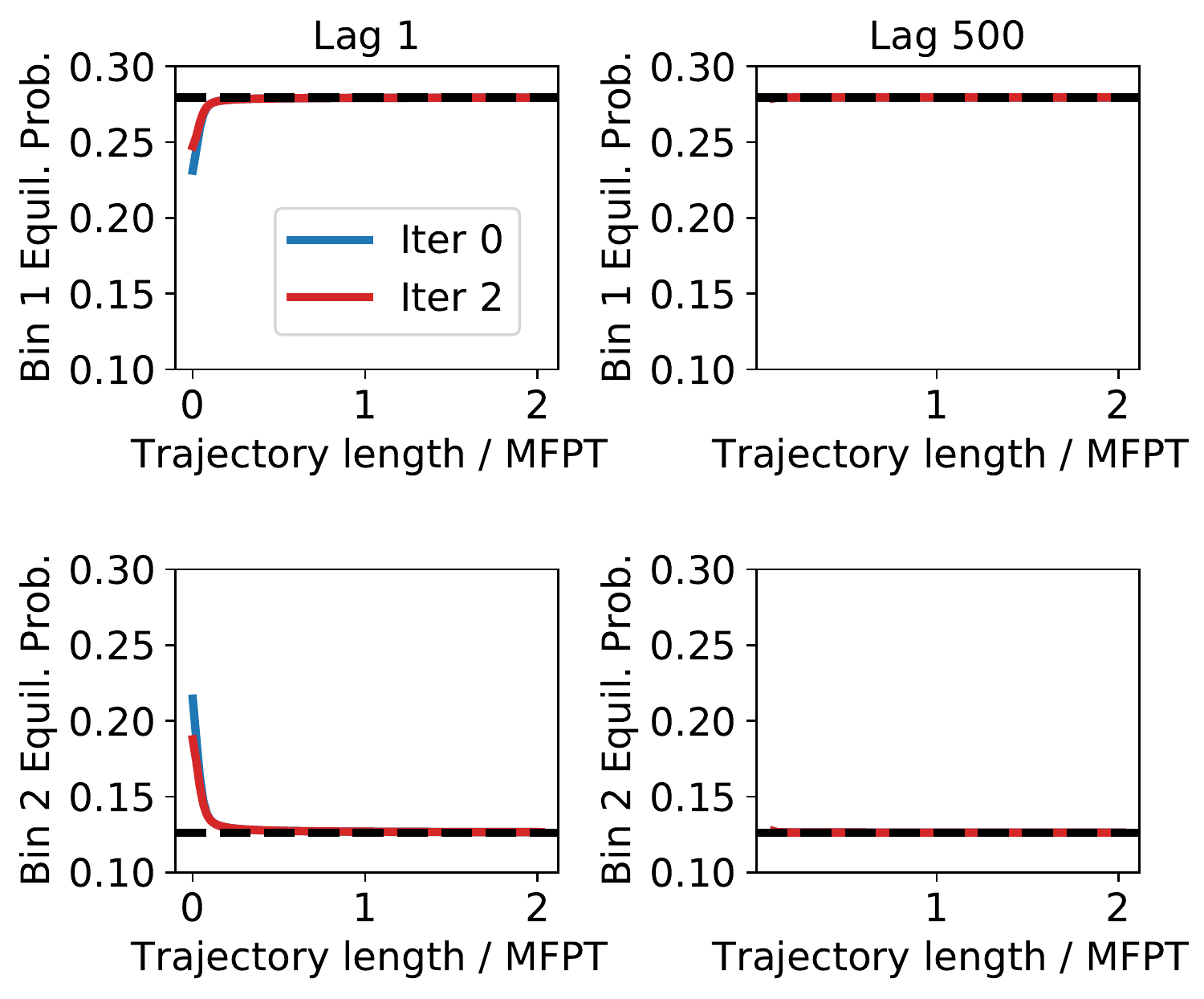}
    \caption{\textbf{Iterative reweighting for equilibrium estimation.} Initial equilibrium estimate (blue lines) and subsequent iterative estimation (2 iterations, red lines), and reference value (black lines).
    Estimates use the MSM stationary distribution based on $\meso(S)$ for the trajectory lengths shown.
    We assumed a non-informative uniform initial distribution of weights $w_i(0) = 1/42$.
    }
    \label{fig:ach_asympt_equil}
\end{figure}

For the equilibrium estimator shown in Fig.~\ref{fig:ach_asympt_equil}, the effect of reweighting is apparent but not qualitatively large. Iterative reweighting improves the initial estimates but does not substantially accelerate the timescale of the convergence in our model system.

\begin{figure}[h]
    \centering
    \includegraphics[width=\linewidth]{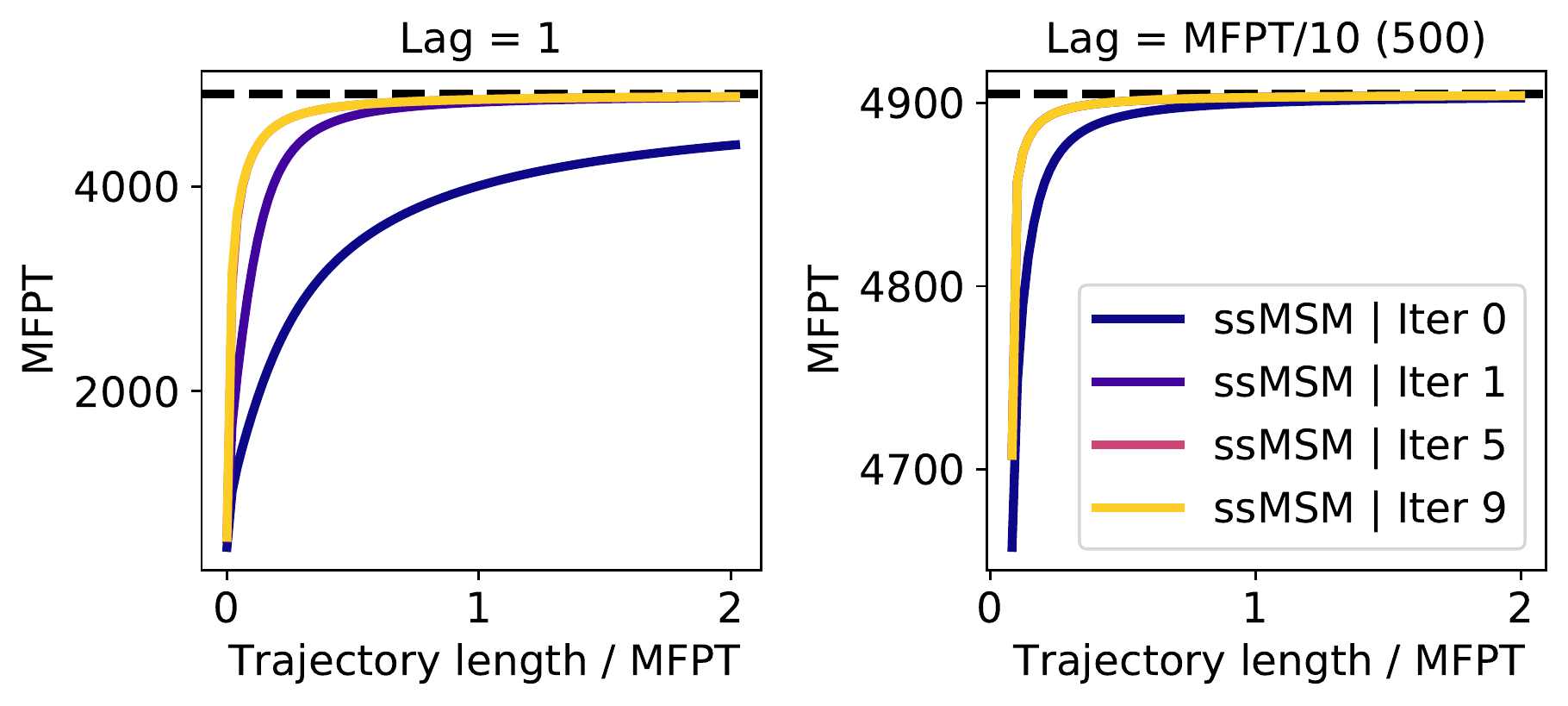}
    \caption{\textbf{Iterative reweighting accelerates $\mathrm{MFPT}$ convergence.} Initial $\mathrm{MFPT}$ estimates (dark blue line) and subsequent iterations (dark purple to orange lines), and reference value (dashed black line).  Estimates are based on the coarse-grained Hill relation \eqref{eq:mfpt-cg} using ssMSMs $\meso^\alpha(S)$ for the trajectory lengths indicated.
    We assumed a non-informative uniform initial distribution of weights $w_i(0) = 1/42$.
    }
    \label{fig:ach_asympt_mfpt}
\end{figure}

\begin{figure}[h]
    \centering
    \includegraphics[width=\linewidth]{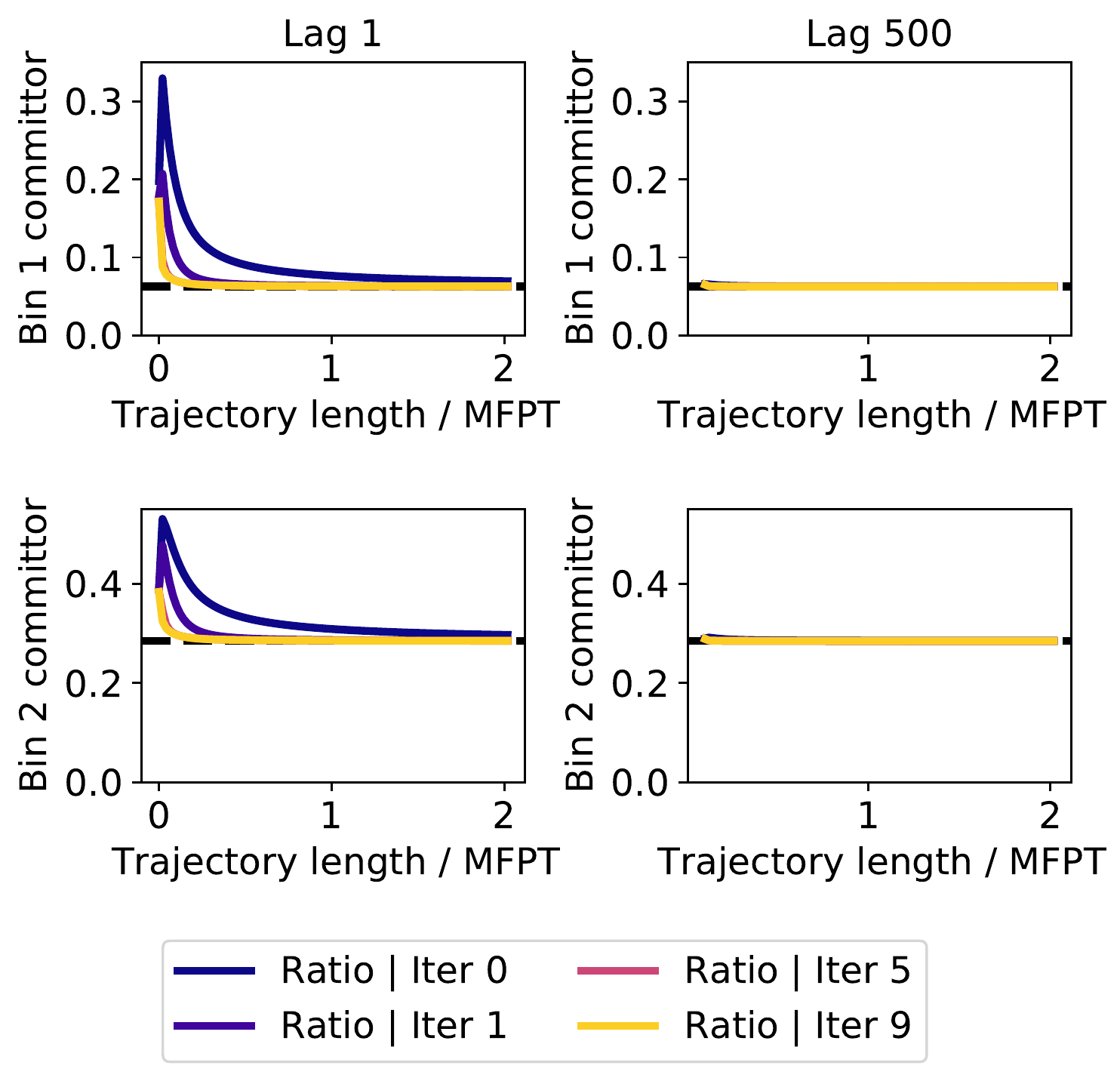}
    \caption{\textbf{Iterative reweighting accelerates committor convergence} Initial committor estimates (dark blue lines) and subsequent iterations (dark purple to orange lines), and reference value (dashed black line).
    Estimates employ the ratio estimator \eqref{eq:qbig} applied to stationary solutions of the ssMSM and MSM at the trajectory lengths indicated.
    We assumed a non-informative uniform initial distribution of weights $w_i(0) = 1/42$.
    }
    \label{fig:ach_asympt_comm}
\end{figure}

Convergence of first-passage times (Fig.~\ref{fig:ach_asympt_mfpt}) and committors (Fig.~\ref{fig:ach_asympt_comm}), are substantially accelerated. 
Using the iterative reweighting approach, the convergence timescale was reduced from multiple first-passage times, to roughly half a first-passage time.
This effect is more pronounced for the short-lag, where the initial state bias affects the estimates more drastically as previously discussed.

\section{Conclusions}
This study has explored unbiased estimation of observables using MSMs (and variants) with a particular focus on convergence with increasing trajectory length. 
Using exact discrete-state calculations enabled us to sidestep sampling concerns.

Although it has been known that standard MSMs in principle provide unbiased estimation of equilibrium populations \cite{SAZ, NCZ} and also that history traceback could allow unbiased estimation of the MFPT \cite{SAZ, NCZ}, we believe that unbiased estimators for the committor values of coarse-grained states were not previously available in a MSM framework.  These estimators highlight the critical importance of boundary conditions (applied before constructing the transition matrix), which was not previously appreciated as far as we know.  Furthermore, the relaxation properties of the estimators were not previously assessed to our knowledge.  We emphasize that relaxation of estimated observables (as sliding-window averaging occurs) is not an abstract issue, but directly impacts whether unbiased estimates can be obtained using feasible amounts of data and trajectory lengths.  We also showed that extending the reweighting idea proposed by Voelz and coworkers \cite{Voelz} has the potential to make a significant difference in practical unbiased estimation.  Although our work relied on a discretized microscopic dynamics, it is not difficult to see that almost identical considerations apply to continuous trajectories.

\begin{acknowledgments}
We gratefully acknowledge support from the National Institutes of Health via Grant GM115805 and from the National Science Foundation via Grant DMS-181871 to DA and GS.  Early discussions with Ernesto Suarez were of great value.  

\end{acknowledgments}

\FloatBarrier

\bibliography{bibliography}

\clearpage\FloatBarrier
\onecolumngrid
\appendix
\appendixpage
\counterwithin{figure}{section}
\section{Microscopic transition matrix}
\label{app:matrix}

\begin{figure}[ht]
    \centering
    \includegraphics{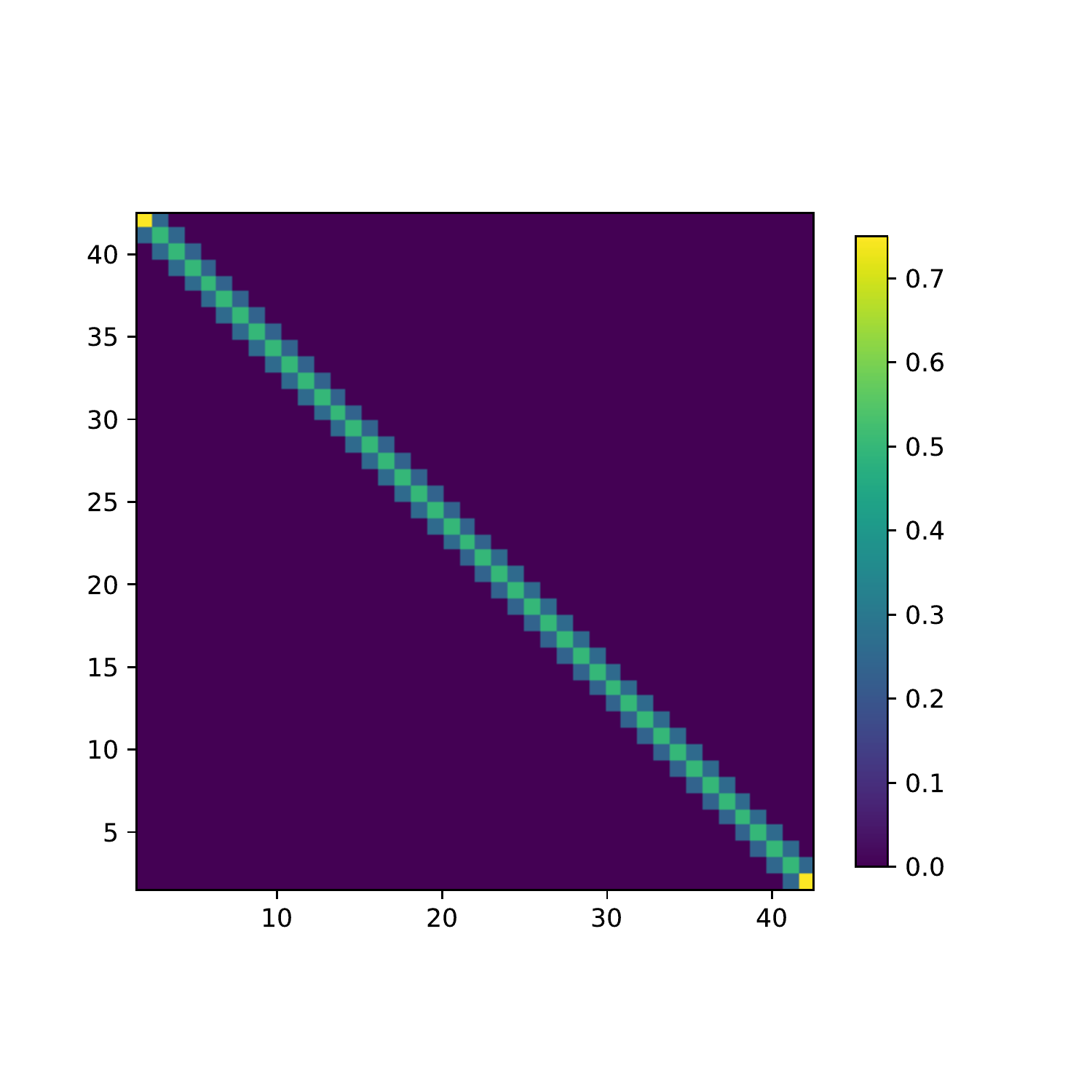}
    \caption{Heatmap of the microscopic transition matrix $\micro$. 
    Microstates at the left and right boundary (i.e. microstates 1 and 42) have a 0.75 self-transition probability, and 0.25 transition probability to the neighbor. 
    Microstates 2 and 41 have a 0.5 self-transition probability, and 0.25 transition probability to each neighbor. 
    A barrier is introduced by giving all other microstates a 0.5 self-transition probability, a 0.24 transition probability to the adjacent microstate closer to the middle of the system, and 0.26 transition probability to the adjacent microstate away from the middle.}
    \label{fig:p}
\end{figure}

\end{document}